\documentstyle[draft,psfig]{mn}
\def\gs{\mathrel{\raise0.35ex\hbox{$\scriptstyle >$}\kern-0.6em
\lower0.40ex\hbox{{$\scriptstyle \sim$}}}}
\def\ls{\mathrel{\raise0.35ex\hbox{$\scriptstyle <$}\kern-0.6em
\lower0.40ex\hbox{{$\scriptstyle \sim$}}}}

\date{\fbox{\sc Draft: \today\ --- Do Not Distribute}}

\title[Lensing in blank-field SCUBA surveys]{
The effect of lensing on the identification
of bright SCUBA galaxies.}

\author[Chapman et al.]
{S.\,C.\ Chapman,$^{\! 1}$
Ian Smail,$^{\! 2}$
R.\,J.\ Ivison$^{3}$ \&
A.\,W.\ Blain$^1$
\vspace*{1mm}\\
	$^1$ California Institute of Technology, Pasadena, CA 91125\\
        $^2$ Department of Physics, University of Durham, South Road,
        Durham DH1 3LE\\
        $^3$ Astronomy Technology Centre, Royal Observatory, Blackford Hill,
        Edinburgh EH9 3HJ}

\date{Accepted ... ; Received ... ; in original form ...}

\pagerange{000--000}

\begin{document}

\maketitle

\begin{abstract}
Spectroscopic surveys of luminous submillimetre-selected sources have
uncovered optically-bright galaxies at $z\ls 1$ close to the positions
of several submillimetre (submm) sources.  Naive statistical analyses
suggest that these galaxies are associated with the submm emission.
However, in some cases, it is difficult to understand this association
given the relatively modest redshifts and unpreposessing spectral
characteristics of the galaxies. These are in stark constrast to those
expected from the massive dust-enshrouded starbursts and AGN thought
to power the bulk of the bright submm population.  We present new
observations of optically-bright counterparts to two luminous submm
sources, along with a compilation of previously proposed
optically-bright counterparts with $z\ls 1$.  We suggest that the
majority of these associations between bright galaxies and submm
sources may be due to the action of the foreground galaxies as
gravitational lenses on the much fainter and more distant submm
sources.  We discuss the implications of this conclusion for our
understanding of the SCUBA population.
\end{abstract}

\begin{keywords}
   galaxies: starburst
-- galaxies: formation 
-- infrared: galaxies
\end{keywords}

\section{Introduction}

%
%
\begin{figure*}
\centerline{\psfig{figure=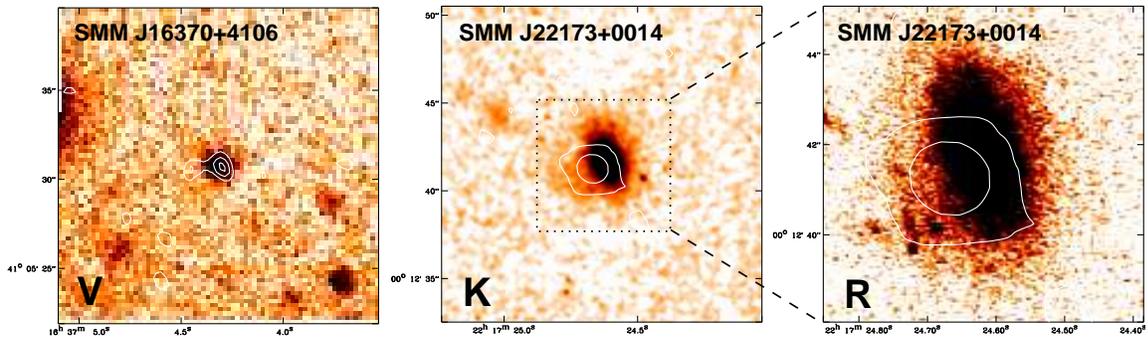,angle=0,width=6.0in}}
\caption{\footnotesize
Images (18\arcsec$\times$18\arcsec) of the two $z<1$ candidate submm
sources from our ESI survey: SMM\,J16370+4106 (left panel, $V$-band)
and SMM\,J22173+0014 (centre panel, $K$-band).  The right-hand panel
shows a zoomed $7''\times 7''$ view of SMM\,J22173+0014 as seen in an
{\it HST} STIS $R'_{573}$ image.  Radio contours are overlaid starting
at $3\sigma$, and increasing in $1\sigma$ intervals.
}
\label{fig1}
\end{figure*}

The intentional use of gravitational lenses has been a particularly
profitable route to probe the nature of the submm galaxy population
(Smail, Ivison \& Blain 1997; Chapman et al.\ 2002a; Smail et al.\
2002; Cowie, Barger \& Kneib 2002).  The lens amplification
facilitates the identification and characterisation of counterparts at
other wavelengths (Smail et al.\ 2002), and has produced some of the
best-studied examples of the submm population (e.g.\ Ivison et al.\
1998, 2000; Soucail et al.\ 1999).  Blank field submm surveys have
also been undertaken (e.g.\ Hughes et al.\ 1998; Barger et al.\ 1999;
Eales et al.\ 1999, 2000; Borys et al.\ 2002; Scott et al.\ 2002; Webb
et al.\ 2002), these cover large contiguous areas and are in principle
simpler to analyse than the lens surveys.  However, without the boost
from a cluster lens, identifying counterparts in the blank field
surveys has proved to be an arduous task.  For example, significant
effort has been expended in attempting to track down the brightest
submm source in the Hubble Deep Field North, HDF\,850.1, with no clear
resolution yet.

The published blank field SCUBA surveys have reliably identified a
small fraction of their submm sources in the optical and near-infrared
(Gear et al.\ 2000; Lutz et al.\ 2001), with a mere handful having
spectroscopic redshifts, e.g.\ Westphal-MMD11 (Chapman et al.\ 2002b);
CUDSS\,3.8, 3.10, 14.13 and 14.18 (Eales et al.\ 2000; Webb et al.\
2002).  Although the spectral properties of some of these proposed
counterparts appear reasonable given their redshifts, several of them
must have very low dust temperatures compared to similar luminosity
galaxies in the local Universe, if they are to produce the observed
submm emission at the proposed redshifts, $z\ls 1$ (Eales et al.\
1999; Dunne \& Eales 2001).

There are only three SCUBA galaxies with spectroscopic redshifts which
have been confirmed in CO line emission, all these lie at $z\geq 1$
(e.g.\ Frayer et al.\ 1998, 1999), consistent with the estimates of
the median redshift for the whole population of $z\sim 2$--3 from
their submm and radio spectral properties (Hughes et al.\ 1998;
Carilli \& Yun 2000; Smail et al.\ 2000; Yun \& Carilli 2002), with
few galaxies at $z\ll 1$.  The existence of a moderate fraction of
SCUBA galaxies with redshifts of $z\ll 1$ would therefore indicate a
bimodal redshift distribution for the population, and suggest that two
physically distinct classes of SCUBA galaxies contribute to the faint
submm counts.  Alternatively, these apparent low redshift counterparts
could be misidentifications -- but why are they so frequent?

Given the expected high median redshift of the submm population and
the steep number counts seen for the brighter submm sources (Blain et
al.\ 1999; Fox et al.\ 2002), gravitational lensing by foreground
galaxies is one possible explanation for the anomolously high rate of
association of submm sources with low redshift, optically bright
galaxies.  Surveys of distant QSOs have provided estimates of the rate
of lensing for optical samples of $z>1$ QSOs, 0.7\% (e.g.\ Surdej et
al.\ 1993; Jaunsen et al.\ 1995; Kochanek et al.\ 1995, see also
Kochanek 1993).  A more reliable constraint comes from lens searches
in the radio waveband where obscuration in the lens is not a concern,
these suggest that a modest fraction of flux limited radio-loud AGN
samples, 0.25--0.57\%, are strongly lensed by field galaxies at $z\ls
1$ (e.g.\ King et al.\ 1999; Myers et al.\ 1999).  This same
population of lenses will also act on the similarly distant SCUBA
population, although differences in the intrinsic count slope of the
two populations will lead to a different rate of occurence of strong
(and weak) lensing in flux limited submm samples.  The issue of the
effects of gravitational lensing on field surveys in the submm
waveband have been discussed by Blain (1996; 1998), who suggested that
$\sim 2$\% of SCUBA galaxies with S$_{\rm 850 \mu m}\simeq 10$\,mJy
will be amplified by $\geq 2\times$ by foreground lenses.  The number
of submm sources in published samples is now sufficient that we should
be able to observationally test this prediction and in doing so
investigate the nature of those SCUBA galaxies which apparently lie at
$z\ll 1$.

\section{Observations and Reduction}

We have embarked on a program to obtain spectroscopic redshifts for a
large sample of radio-selected, optically bright submm sources.  The
goal of this project is to identify the optically-brightest SCUBA
galaxies, such as SMM\,J02399$-$0136 and SMM\,J14011+0252 (Ivison et
al.\ 1998; 2000, 2001) which are suitable for detailed follow-up on
10-m class telescopes.  To achieve this we extend the previous work on
submm observations of optically-faint radio-selected sources (e.g.\
Chapman et al.\ 2001b) to brighter counterparts.  Chapman et al.\
(2002e) have demonstrated the success of this approach, recovering
$\sim$60\% of the SCUBA population.  The expectation is that the
remaining 40\% include both optically luminous SCUBA galaxies (Ivison
et al.\ 1998, 2000) and very high redshift sources which are too faint
in the radio waveband to be detected (e.g.\ Frayer et al.\ 2000).  The
bias against very distant SCUBA galaxies in radio surveys depends on
their dust temperature: hotter sources can be seen out to higher
redshifts, $z\gs 3$, while colder sources are more difficult to detect
(Chapman et al.\ 2002d).

%
%
\centerline{\psfig{figure=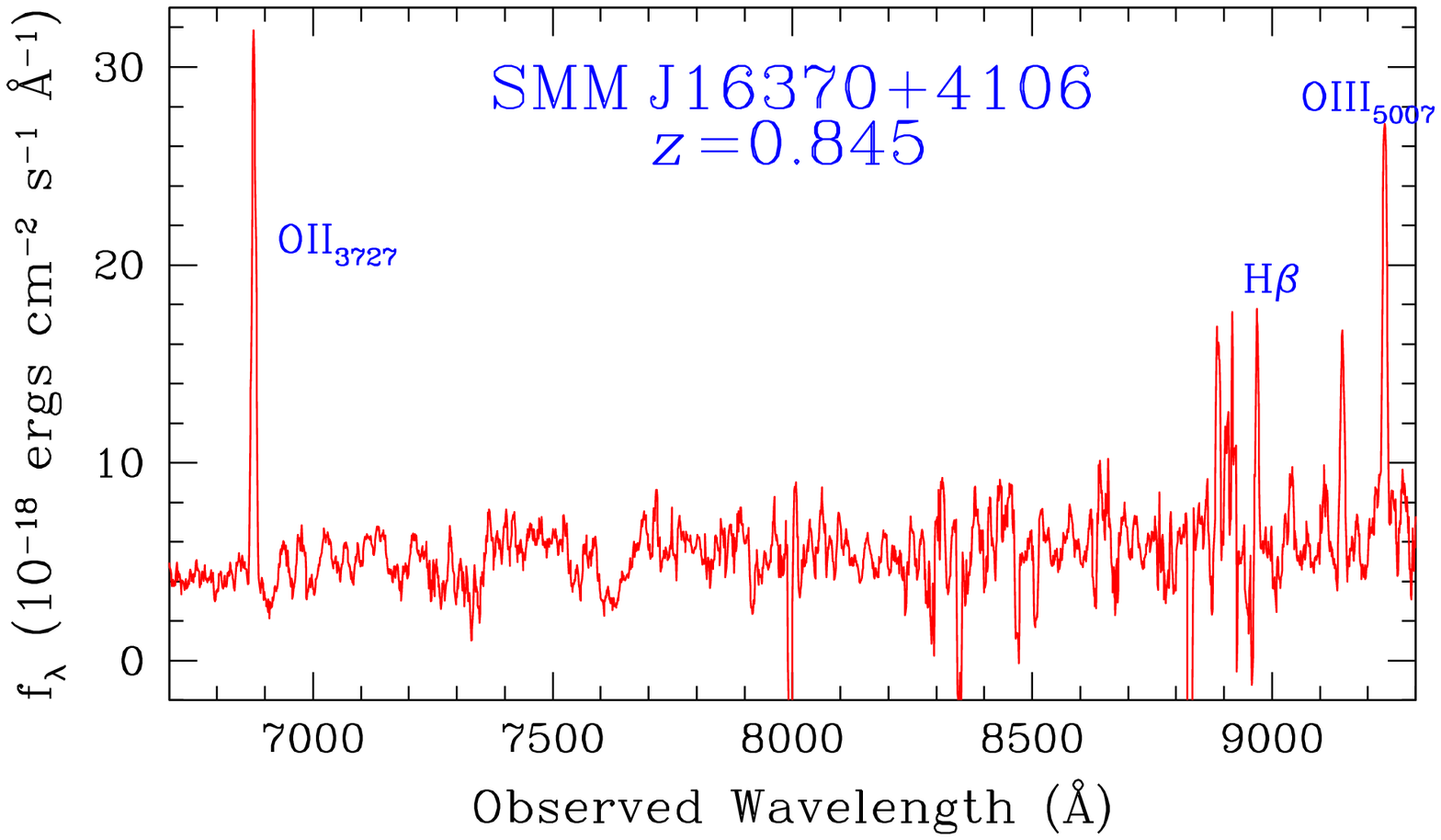,angle=0,width=2.5in}}
\centerline{\psfig{figure=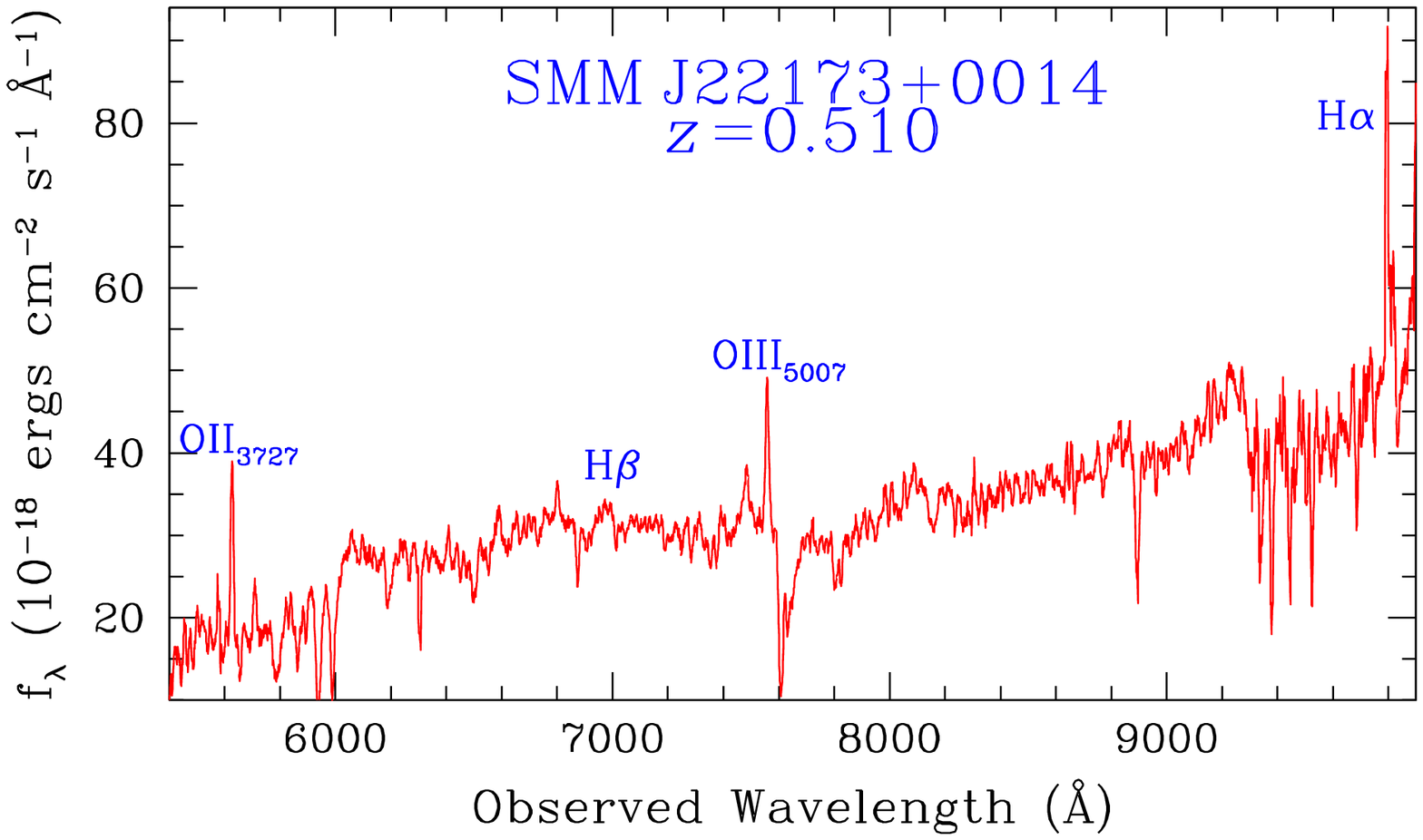,angle=0,width=2.5in}}
\noindent{\footnotesize \addtolength{\baselineskip}{-3pt}
{\bf Fig.~2:}  Keck/ESI spectra of SMM\,J16370+4106 (upper
panel) and SMM\,J22173+0014 (lower panel). We identify redshifted
emission lines in the spectra of both galaxies which place them at
$z<1$.

}

\subsection{ESI Observations}

The initial observations for our survey of optically bright SCUBA
sources involved spectroscopy of 7 sources with the Echellette
Spectrograph and Imager (ESI) on the Keck\,II telescope in 2001 July.
The echellette mode provides complete coverage from 0.32--1\,$\mu$m at
high spectral resolution $\sim$1\AA, allowing very good sky
subtraction into the atmospheric OH forest at the red end of the
spectra.  The spectral integrations were 1800\,s for each source.
High signal-to-noise flats and wavelength calibrations were taken
shortly before the observations of each target, and fluxes were
calibrated using spectra of red standard stars.  The data were reduced
with the {\sc makee} software using the reduction recipe described in
Barlow \& Sargent (1997).

%
%
\begin{table*}
{\footnotesize
\centerline{\sc Properties of Sample}
\begin{center}
\begin{tabular}{lcccccccl}
\hline\hline
\noalign{\smallskip}
 {Source}  & $\delta r^a$ &  $z_{\rm lens}$ & $z_{\rm CY}$ & T$_{\rm d}^b$ &  
   $I$-mag$^c$ &  $S_{850 \rm \mu m}$ & $S_{1.4 \rm GHz}$ & \cr
     {}    &  $('')$    &                 &              & (K)         &     &   (mJy)              & ($\mu$Jy)         & \cr
\noalign{\smallskip}
\hline
\noalign{\smallskip}
 SMM\,J16370+4106 & 2.0 & ~0.845 &2.2 
& 23 & 21.4 &  11.2$\pm$2.0 & 99$\pm$20 & N2\,850.1, Ivison et al.\ (2002) \cr
 SMM\,J22173+0014 & 1.7 & ~0.510 &2.4
& 19 & 19.4 &  15.0$\pm$3.0 & 145$\pm$18 & SSA22\,850.2, Chapman et al.\ (2001a) \cr
\noalign{\medskip}
 SMM\,J00266+1708$^d$ & 1.5 & ~0.44 & 2.9 & 16 & 22.0 &  18.6$\pm$1.5 & 100$\pm$15 & ERO: M12, Frayer et al.\ (2000) \cr
 SMM\,J04431+0210$^d$ & 2.3 & ~0.18 & $>2.6$~ & 14 & 18.4 &  7.2$\pm$1.5 & $<57$, 3$\sigma$ & ERO: N4, Smail et al.\ (1999) \cr
 SMM\,J12369+6212 & 0.7 & $\sim$1.1 & 4.1 & 18 & 22.6 & 7.4$\pm$0.5 & $<24$, 3$\sigma$ & HDF\,850.1, Downes et al.\ (1999) \cr
\hline
\end{tabular}

\begin{tabular}{l} 
$^a$ Angular separation between the bright galaxy and the
nominal position of the submm, mm or radio centroid.\cr
$^b$  T$_{\rm d}$ is the dust temperature required for the source to
have a radio/submm redshift consistent with the spectroscopic \cr
~~ measurement of the galaxy.  \cr
$^c$  The $I$-band magnitude of the optically bright counterpart. \cr
$^d$  Fluxes have not been corrected for cluster lensing. \cr
\end{tabular}

\end{center}
}
\label{tab1}
\end{table*}

These observations generated redshifts for five new candidate
counterparts to submm galaxies, to be discussed in a future paper.
However, the proposed optically-bright counterparts to two of these
sources lie at much lower redshifts than expected from their radio and
submm properties, and we discuss these further here.  These two
sources are: SMM\,J16370+4106 from the SCUBA survey of Scott et al.\
(2002) and SMM\,J22173+0014 from Chapman et al.\ (2001a).  We show
radio/optical overlays of these two sources in Fig.~1 and summarise
their properties in Table~1.  The calibrated spectra are shown in
Fig.~2 with their spectral features identified.  These spectra show
that the proposed optically-bright counterpart to SMM\,J16370+4106
lies at $z=0.845$, while the galaxy identified with SMM\,J22173+0014
has $z=0.510$.  We note that the spectrum of SMM\,J22173+0014 shows
broad lines indicative of a low luminosity type~I AGN, although such
activity is not infrequent in the high-redshift field population
(Cowie et al.\ 1996).

As we discuss below the properties of both of these systems strongly
suggest that the optically bright, spectroscopically-identified
galaxies are not in fact the source of the submm emission, but instead
they may represent foreground gravitational lenses which are
amplifying the more distant, optically-faint SCUBA galaxy.

\section{Analysis and Discussion}

We have cataloged two new, $>10$\,mJy submm sources which are within
1--2$''$ of $I<21.5$ galaxies at $z=0.510$ and $0.845$.  In the
absence of detections in deep, high resolution radio maps, the coarse
spatial resolution of the SCUBA detections would mean that these
offsets would not be significant, and combined with the low surface
density of $I<21.5$ galaxies this would lead to a high probability
that the submm source is associated with the optically bright
galaxy. There are two possible explanations for these associations:
1) either the optically-bright galaxies are the source of the submm
emission, or 2) these galaxies are lensing the true submm source,
which is then both fainter and lies at much higher redshifts.

If the spectroscopically-identified galaxies are the source of the
submm and radio emission then we can estimate the characteristic dust
temperatures, T$_{\rm d}$, which these galaxies must have to be
consistent with the measured radio/submm spectral index. We list these
values in Table~1 and compare them to the distribution of dust
temperatures (Dunne et al.\ 2000) as a function of galaxy luminosity
seen locally in Fig.~3. The distribution is shown as log normal,
consistent with the best fit distribution for the local 1.2\,Jy IRAS
sample (Chapman et al.~2002f).

In the case that these systems represent gravitational lenses we list
for comparison in Table~1 the properties of three sources from
published SCUBA surveys which are all believed to be lensed by
foreground galaxies: SMM\,J00266+1708 (Frayer et al.\ 2000) and
SMM\,J04431+0210 (Smail et al.\ 1999), both of these lie within $\sim
2''$, but are not coincident with, bright foreground galaxies (images
of these systems can be found in the relevant references).  We also
include the candidate lensed source, HDF\,850.1 from Hughes et al.\
(1998).  The millimetre interferometry map of HDF\,850.1 by Downes et
al.\ (1999) indicates that the submm source is offset by 0.7\arcsec\
from a moderately bright Elliptical galaxy at modest redshift,
suggesting it is likely to be lensed (Downes et al.\ 1999).  
Very recently an extremely faint $K$-band source has been
found at the millimetre position, providing further support for the
identification of this system as a high redshift lensed SCUBA galaxy
(Dunlop et al.\ 2002).

We now discuss the properties of the two new submm sources to attempt
to distinguish between the two possible explanations of their
properties.

For SMM\,J16370+4106 we find a close pair of radio sources -- the
brighter radio component is consistent with being coincident with the
optically-bright galaxy, within the optical/radio astrometric
uncertainty ($\sim0.2$\arcsec).  The second, fainter radio component
lies $\sim 2''$ east of the bright optical galaxy.  The two radio
components could both be associated with the bright galaxy as the
chance of a radio/submm source being lensed by another foreground
radio source is low, due to the low radio source density at these flux
levels.  At the same time, the astrometric uncertainty does not allow
us to reject the possibility that this system is a highly amplified
double image lensing configuration, straddling the optical lens, which
would circumvent the radio source number density/probability argument.
The required T$_{\rm d}$ for SMM\,J16370+4106 at $z=0.845$ is $23\pm
5$\,K, placing it $\sim 4\sigma$ below the local relation in Fig.~3
and strengthening the lensing hypothesis.
         
SMM\,J22173+0014 lies only 30\arcsec\ from a known submm source at the
centre of a $z=3.1$ galaxy overdensity (Steidel et al.\ 1998), yet
appears to be associated with a luminous early-type spiral at
$z=0.510$. However, the radio centroid is significantly offset from
the optical galaxy center, five times the relative astrometric error
in the VLA/optical frames, which is dominated by the source
centroiding, ${\rm FWHM}/(2\times S/N) = {5\arcsec}/{17} =
0.3\arcsec$, with the radio emission falling close to a faint $K$-band
extension from the galaxy. The {\it HST} image obtained subsequent to
the spectroscopic observations reveals a morphologically complex
systems, reminiscent of a merger, coincident with this extension.  In
addition, the calculated T$_{\rm d}$ required to reproduce the
radio/submm spectral index at this redshift is only $18\pm 2$\,K,
putting this source $\sim 5\sigma$ below the local mean T$_d$ for its
luminosity (Fig.~3).  Together, these arguments suggest that
SMM\,J22173+0014 is a distant submm source lensed by the foreground
galaxy at $z=0.510$.  The configuration of this system is very similar
to that seen in SMM\,J00266+1708 and SMM\,J04431+0210, which have
bright edge-on spiral galaxies within 2--3$''$ of the submm source
(Frayer et al.\ 2000; Smail et al.\ 1999).  These configurations are
also consistent with the results from Blain, M\"oller \& Maller (1999)
and Bartelmann \& Loeb (1998), who have shown that nearly edge-on
spirals provide a high lensing probability, due to the large projected
surface mass density.

%
%

\centerline{\psfig{figure=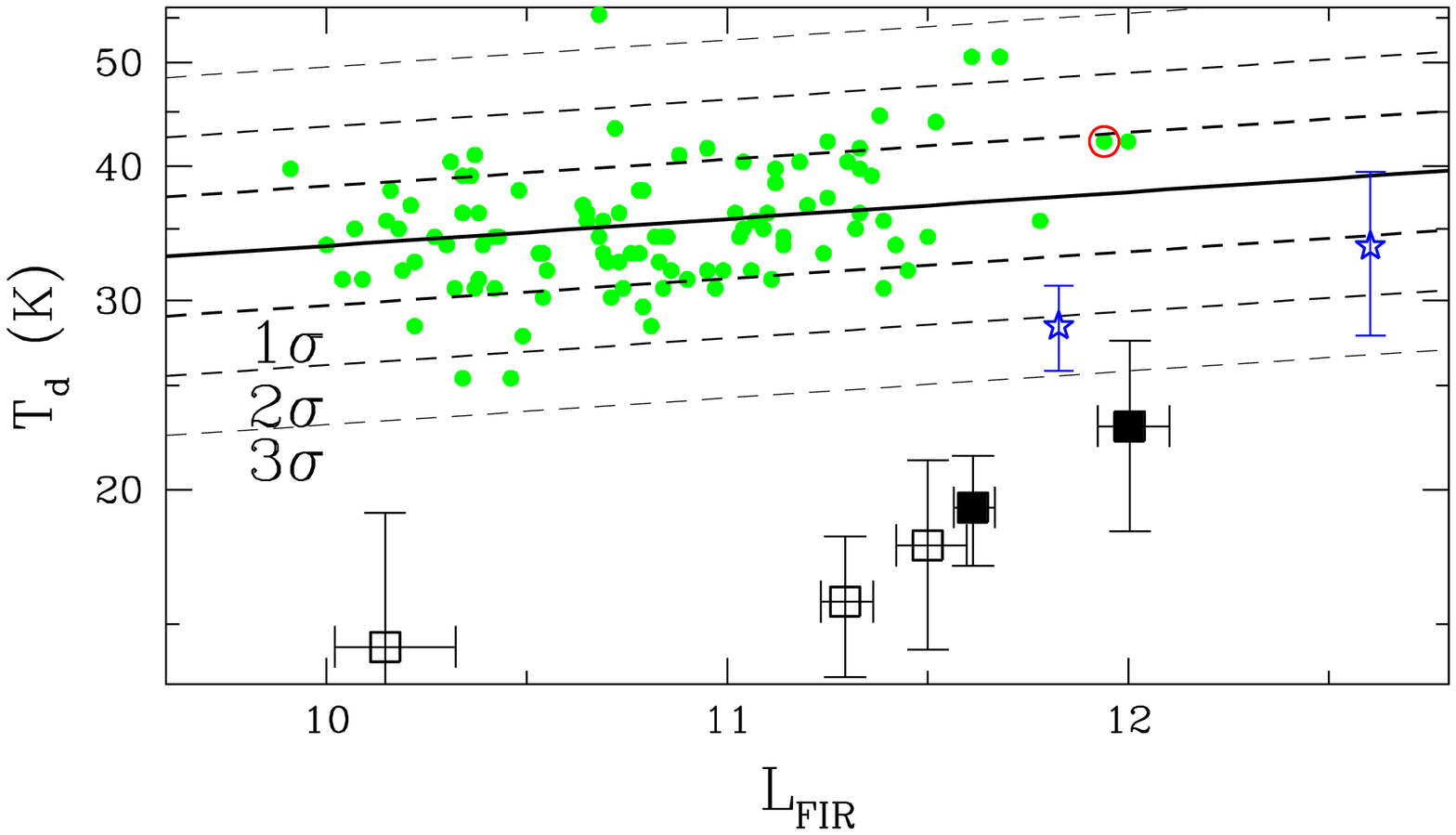,angle=0,width=3.5in}}
\noindent{\footnotesize\addtolength{\baselineskip}{-3pt}
{\bf Fig.~3:} The distribution of T$_{\rm d}$ versus far-infrared
luminosity for local luminous, dusty galaxies (circles), the best fit
relationship is shown as a solid line and the error bounds by dashed
lines.  We show our two new systems by filled squares, assuming that
the submm and radio emission is associated with the bright optical
galaxy, as well as the lensed submm sources from the literature (open
squares).  We also show two distant, luminous {\it ISO}-FIRBACK
galaxies from Chapman et al.\ (2002d) which have low dust temperatures
(stars).

}

While there is strong evidence that the two new submm sources are both
lensed systems, we first need to consider selection effects in the
radio/submm sample.  The local luminous {\it IRAS} galaxies in Fig.~3
show a broad distribution in T$_{\rm d}$, with no examples as cold as
the SCUBA galaxies would need to be to be $z<1$.  However, the {\it
IRAS} galaxies represent a restframe 60-$\mu$m selected sample, while
the SCUBA galaxies are initially identified in terms of their 1.4-GHz
and 850-$\mu$m emission.  This will tend to bias their selection
towards galaxies with lower T$_{\rm d}$ (Eales et al.\ 1999).  This
effect has recently been demonstrated for two SCUBA sources from the
{\it ISO}-FIRBACK survey (Chapman et al.\ 2002d).  These 170-$\mu$m
selected sources have accurate radio positions which identify them
with relatively low redshift galaxies ($z=$0.45 and 0.91) which both
exhibit clear merger morphologies. The best fit dust temperatures for
these luminous galaxies (calculated for consistancy with the sample in
Fig.~3) are 28\,K and 33\,K, indicating that some cold, but luminous
galaxies do exist at high redshifts (Fig.~3).  There is also some
evidence that the dust temperature distribution may be broader at
higher redshifts (Chapman et al.\ 2002f), making these apparently cold
sources more common.  The 1-$\sigma$ errors on SMM\,J16370+4106
overlap with those of the cold FIRBACK sources, suggesting that the
cold dust explanation is conceivable for this source.  However, none
of these galaxies have as low T$_{\rm d}$ as is required for
SMM\,J22173+0014 and we therefore conclude that SMM\,J22173+0014 most
probably represents a lensed SCUBA galaxy.
 
Assuming that at least one, and perhaps both, of these two submm
sources are likely to be lensed we now estimate the proportion of
similarly lensed sources in a typical SCUBA survey.  Our survey
started with a subsample of the total submm population which were
radio identified, with the requirement that the radio source aligned
with an $I<23.5$ optical galaxy to within 2\arcsec.  The fraction of
submm sources in total which have a bright ($I<23.5$) galaxy nearby
(Chapman et al.\ 2002c) represents 16\% of radio identified submm
population.  Radio selection detects $\sim60$\% of all bright
($S>5$\,mJy) submm galaxies.  We found above that 1--2 from 5 sources
in our Keck/ESI sample were likely to be lensed.  This implies 3--5\%
of the radio detected submm population could be lensing candidates,
with a lower limit of 2--3\% for the whole SCUBA population.

We have also discussed three lensed SCUBA galaxies from the literature
which are very similar to the two candidate systems presented here.
The submm source HDF\,850.1 appears to be lensed by a foreground
elliptical galaxy, this gives a rate of lensed sources of 1/8, or
$\sim 13$\% with a large uncertainty, for the HDF sources discussed by
Hughes et al.\ (1998) and Serjeant et al.\ (2002).  The other two
lensed sources come from the Smail et al.\ (2002) cluster sample,
these are of course more likely to suffer galaxy-galaxy lensing due to
the high foreground galaxy concentration in the clusters used in this
survey.  For that reason we take the rate of galaxy-lensing from this
survey as an upper limit, 2/15 sources, or $\leq 13$\%.  Overall this
suggests that around 3--5\% of submm sources from SCUBA surveys are
likely to be gravitationally amplified by foreground galaxies.

The rate of lensing we find is nearly an order of magnitude higher
than the incidence of multiple-imaging in similarly distant QSO
samples, which combined with the radio morphologies of the galaxies
discussed here indicates that these are not highly-amplified,
multiply-imaged SCUBA sources.  With our current crude knowledge of
the lensing configurations in these systems, it is impossible to
produce detailed lens models for these sources.  However, the relative
separation of the source and lens, along with the apparent lack of
multiple images suggests amplifications of $\ls 5$.

The probability of lensing in SCUBA surveys depends on the form of the
submm counts, with the steep count slope seen for submm sources
brighter than 5\,mJy (Scott et al.\ 2002), $\sim2$\% of the
$\sim10$\,mJy sources could be amplified by $\geq 2\times$ (Blain
1998).  This figure is roughly in agreement with our findings,
especially if SMM\,J16370+4106 actually represents a cold, luminous
galaxy.

We note that weak amplification bias similar to, but weaker than, that
invoked here has been invoked as a possible explanation for the strong
cross correlation between foreground bright galaxies and submm sources
seen by Almaini et al.\ (2002).  We also note that lensing is also
expected to manifest at the $\sim 1$\% level given the surface density
of sources now reached in the deepest radio maps, $\sim 15\mu$Jy
(e.g.\ Formalont et al.\ 2002; Owen et al.\ 2002). However, the effect
cannot be much greater as $\mu$Jy count slopes are close to Euclidean,
and the $N(z)$ is less extended than for the submm-selected samples
($<\! z\! > \sim 0.6$, Richards et al.\ 1999; Barger, Cowie \&
Richards 2000; Chapman et al.\ 2002c).

\vspace*{-2mm}
\section{Conclusions}

We have identified two submm sources which have apparent counterparts
which are optically bright galaxies, $I<21.5$, lying at modest
redshifts, $z<1$. We suggest two possible explanation for these
sources, and similar systems in the literature.  The first explanation
states that the suggested identifications are correct, in which case
these galaxies represent a class of very cold, luminous submm source
(Chapman et al.\ 2002d).  For one source, SMM\,J16370+4106 at
$z=0.845$, this interpretation appears possible, however, it seems
less likely for the second and even colder source, SMM\,J22173+0014.
In both cases, the most powerful test to reject this hypothesis would
be to confirm the absence of luminous molecular CO emission at the
redshift of the optically bright galaxies.

The second explanation for these systems is that the optically bright
galaxy represents a foreground gravitational lens, which is amplifying
the more distant SCUBA source.  This explanation is consistent with
both the observed source configurations and their spectral properties.
In this case we estimate that up to 3--5\% of the $\gs 10$\,mJy submm
sources in blank field SCUBA surveys could be gravitationally
amplified by foreground galaxies.  The most important result of this
amplification is that it leads to the misidentification of the submm
source with the nearby bright, and typically low redshift, galaxies.
These misidentifications would produce a false tail of low-redshift,
$z\ll 1$, submm sources in all far-infrared/submillimetre/millimetre
surveys.  This is especially a concern for those surveys where deep,
high resolution data is not available to confirm the correspondance
between the radio (pin-pointing the longer wavelength emission source)
and optical sources with a precision of $\ll 1''$.

\vspace*{-2mm}
\section*{Acknowledgements}
We acknowledge our collaborators on the HST-SCUBA morphology project,
R.\ Windhorst \& E.\ Richards.
SCC acknowledges support from NASA through HST grant 9174.1.
IRS acknowledges support from a Royal Society URF and a
Philip Leverhulme Prize Fellowship.

\end{document}